\documentstyle[sprocl]{article}

\input{psfig}
\def\be{\begin{eqnarray}}
\def\ee{\end{eqnarray}}

\newcommand{\nc}{\newcommand}
\nc{\al}{\alpha}
\nc{\ga}{\gamma}
\nc{\de}{\delta}
\nc{\ep}{\epsilon}
\nc{\ze}{\zeta}
\nc{\et}{\eta}
\renewcommand{\th}{\theta}
\nc{\ka}{\kappa}
\nc{\la}{\lambda}
\nc{\rh}{\rho}
\nc{\si}{\sigma}
\nc{\ta}{\tau}
\nc{\up}{\upsilon}
\nc{\ph}{\phi}
\nc{\ch}{\chi}
\nc{\ps}{\psi}
\nc{\om}{\omega}
\nc{\Ga}{\Gamma}
\nc{\De}{\Delta}
\nc{\La}{\Lambda}
\nc{\Si}{\Sigma}
\nc{\Up}{\Upsilon}
\nc{\Ph}{\Phi}
\nc{\Ps}{\Psi}
\nc{\Om}{\Omega}
\nc{\ptl}{\partial}
\nc{\del}{\nabla}

\begin{document}

\title{AN ANALYTICAL APPROACH TO LATTICE 
       GAUGE-HIGGS MODELS\footnote{Presented by A.~Ritz. To appear in the
        Proceedings of `Strong and Electroweak Matter (SEWM'97)', Eger,
        Hungary, May 1997.}}

\author{ T.S. EVANS, H.F. JONES AND A. RITZ }

\address{Theoretical Physics Group, Blackett Laboratory, \\
         Imperial College, South Kensington, London, SW7 2BZ, U.K.}


\maketitle
\abstracts{We describe an application of the 
linear $\de$-expansion to the calculation of correlation functions
in $SU(2)$-Higgs lattice gauge theory. A significant advantage of the 
technique is that an infinite volume lattice may be used, allowing 
the non-analyticity in certain observables at a phase transition to be
observed directly. We illustrate the approach with a preliminary 
application to the $3D$ $SU(2)$-Higgs model, as the dimensionally
reduced effective theory for the electroweak standard model at
high temperature, and calculate certain gauge invariant observables
near the phase boundaries.}

\section{Introduction}
When considering the static equilibrium properties of a thermal field  theory,
careful Green function matching\cite{klrs1} has shown that many weakly
coupled gauge theories at high temperature may be reliably approximated
by a super-renormalizable effective $3D$ theory, generically a bosonic
gauge-Higgs model. The dynamical degrees of freedom of this  
effective theory then correspond to the
Matsubara zero-modes of the full theory, thus retaining all the infrared 
divergences and necessitating non-perturbative analysis. However, one has
the advantages of having integrated out the thermally massive fermions.
Furthermore, super-renormalizability, and non-triviality of the scalar sector
in $3D$, allow perturbatively exact renormalization group trajectories 
to be determined with a two-loop calculation, thus allowing control
over the continuum extrapolation of lattice results. With the added benefit of
simulations being less numerically expensive in $3D$, recent Monte Carlo
studies of the dimensionally reduced electroweak
standard model, and MSSM, have led to a reasonably complete knowledge
of the phase structure and related static properties\cite{klrs2}, 
including the insight that the
first order symmetry breaking transition ceases to be of first or
second order for Higgs masses above approximately 80 GeV.

Nevertheless, study of the model near the (possibly second order) 
endpoint of the transition line and, in particular, determination of the
spectrum in this regime require large lattices due to the diverging correlation
lengths. Therefore analytic approaches are still desirable, and 
such approaches may still  take advantage of many of the
features of dimensional reduction discussed above. In this talk we
present an analytic approach, the linear $\de$-expansion, and apply
it to the $3D$ $SU(2)$-Higgs model discretized on an infinite lattice.
We present some preliminary results for certain gauge invariant
`order parameter like' variables near the phase boundaries.

\section{The Linear $\de$-Expansion}
The basis of this approach is that when one is interested in calculating
Greens functions for a theory with classical action $S$, 
one can instead consider the following extended action
\be
 S_{\de} & = & S_0(J) + \de (S-S_0(J)),
\ee
where $\de$ is an artificial parameter.
$S_{\de=1}=S$, the theory under study, and $S_{\de=0}=S_0$ is a 
new `free' action, chosen for convenience of calculation, but also
to mimic the dominant degrees of freedom as closely as possible. 
The generating functional for Green functions 
may then be expanded to an appropriate
order in the parameter $\de$, which is then set to unity. 

Calculating
an observable quantity $P$ at each order in $\de$ 
one then obtains a sequence of approximants $\{P_1,P_2,\ldots,P_N\}$.
An essential feature of this approach is the ability
to optimize the convergence of this sequence, independent of the existence
of a small coupling constant, by carefully fixing additional parameters
$J$ appearing in the extended action.
As the power series in $\de$ is only calculated to a finite order, it retains
some dependence on $J$ which would be absent in a full summation. A well
motivated criterion for fixing $J$ is the principle of minimal
sensitivity (PMS) \cite{stevenson81}, whereby $J$ is chosen at a local
minimum of an observable quantity. i.e. if $P_N(J)$ denotes the $N$'th
approximation to $P$, then we impose
\be
 \frac{\ptl P_N(J)}{\ptl J} & = & 0. \label{pms}
\ee

This or a similar criterion is intrinsic to the success of the 
$\de$--expansion,
providing a generically non-perturbative dependence on the coupling constant,
and there exist rigorous proofs for convergence of the sequence
of approximants for the energy levels $\{E_N\}$, and the partition function,
in 0- and 1-D field theories \cite{converge} 
where the corresponding perturbative series
are asymptotic and eventually diverge factorially or in some cases
are even non-Borel-summable. 

Motivation for the present application to gauge-Higgs models
has come from previous successful studies of pure gauge theories on the
lattice. In particular order parameters \cite{vce}, and
the glueball mass gap \cite{akeyo95} have been well reproduced, while
$4D$ considerations
of the phase structure of the $SU(2)$-Higgs model \cite{zheng91}
have also shown encouraging results.

\section{Application to the $3D$ $SU(2)$-Higgs Model}
Dimensional reduction of a general class of  $4D$ gauge theories at finite 
temperature by static Green function matching has been studied in detail
by Kajantie et al.\cite{klrs1} Perturbative Green function matching
to order $\de G/G\sim O(g^3)$ is possible with a
super-renormalizable effective  $3D$ theory. For the standard model with
Weinberg angle $\th_W=0$, this is the $3D$ $SU(2)$-Higgs model. 

In the present case we can make use of the 
advantages of dimensional reduction discussed earlier, and for reasons
of gauge invariance it is also convenient to use a lattice
regularization. Therefore we consider the standard $3D$ lattice action
\be
 S & = & \beta\sum_P \frac{1}{2}{\rm Tr}U_P + \frac{1}{2} \beta_h \sum_{l_{ij}}
        \rh_i\rh_j {\rm Tr} U_{ij}
         -\sum_i \left[\rh_i^2+\beta_R(\rh_i^2-1)^2\right], \label{latt}
\ee
where we represent the Higgs doublet in the form $\Ph=\rh V$ with
$\rh\in\;${\bf R}$_{+}$ and $V\in Fund[SU(2)]$. We have performed
a gauge transformation of the form $U(x,y)\rightarrow 
V(x)U(x,y)V(y)^{\dagger}$ to write the action in unitary gauge.
The relation between the dimensionless lattice parameters
$\beta,\beta_h,\beta_R$ and the continuum physics, which may be described in
terms of two dimensionless ratios with the scale set by the gauge coupling,
is given in Ref.~2. Super-renormalizability
implies that the necessary renormalization group trajectories may be
obtained exactly with a two-loop perturbative calculation.

For calculation using the linear $\de$-expansion we use the 
following ``free'' action,
\be
 S_0 & = & \sum_{l_{ij}}(J+\beta_h
        \rh_i\rh_j) \frac{1}{2}{\rm Tr} U_{ij}
         -\sum_i \left[\rh_i^2+\beta_R(\rh_i^2-1)^2\right],
\ee
which differs from the full action only in the pure gauge sector,
where a single link action is used which has been successfully applied 
in the analysis of pure gauge theories \cite{vce}. Note that while
formally gauge variant, only gauge invariant expectation values are
nonzero when one allows the parameter $J$ to be site dependent\cite{tan89}.

Expectation values are calculated using the extended action up to a given
order in $\de$. For an observable $P$, the contribution to $N^{th}$ order
has the form
\be
 <P>_N & = & \left.\left[\frac{\sum_{n=0}^{\infty}\frac{\de^n}{n!}
       \int [dU][d\rh\rh^3]P(S-S_0)^ne^{S_0}}{\sum_{n=0}^{\infty}
         \frac{\de^n}{n!}
       \int [dU][d\rh\rh^3](S-S_0)^ne^{S_0}}\right]\right|_{\de^N},
\ee
where the notation implies expansion to $O(\de^N)$. The expectation value then 
assumes the form of a cumulant since the expansion 
of the denominator naturally subtracts the disconnected pieces. 
We present here some preliminary results for the
average plaquette, $E_P\equiv \sum_P {\rm Tr} U_p/2N_P$,
and the hopping term, $E_L \equiv \sum_{<ij>}\rh_i\rh_j {\rm Tr} U_{ij}/2N_L$,
standard local observables which exhibit discontinuities at a transition
point. We assume that the modulus of the Higgs field 
varies slowly over the lattice,
i.e. $\rh_i\rh_j = \rh_i^2+\rh_i\De\rh_i \approx \rh_i^2+O(\de^2)$, 
and then calculations are simplified by noting that all
terms in the action depend either on the real variable $\rh$ or are
$SU(2)$ characters. Using character expansion techniques, one may
reduce all expectation values to polynomials in the factors
$A^n_r/A_0^0$ where
\be
 (A_r^n)^d & = & \int_0^{\infty} d\rh \rh^{n+3}e^{-V_H}
       \left((r+1)\frac{I_{r+1}(J+\beta_h\rh^2)}{J+\beta_h\rh^2}\right)^d,
\ee
in which $V_H=\rh^2+\beta_R(\rh^2-1)^2$ is the Higgs potential,
$I_{r+1}$ is a modified Bessel function, 
and $d$ is the dimension of the lattice.

\begin{figure}
 \centerline{%
   \psfig{file=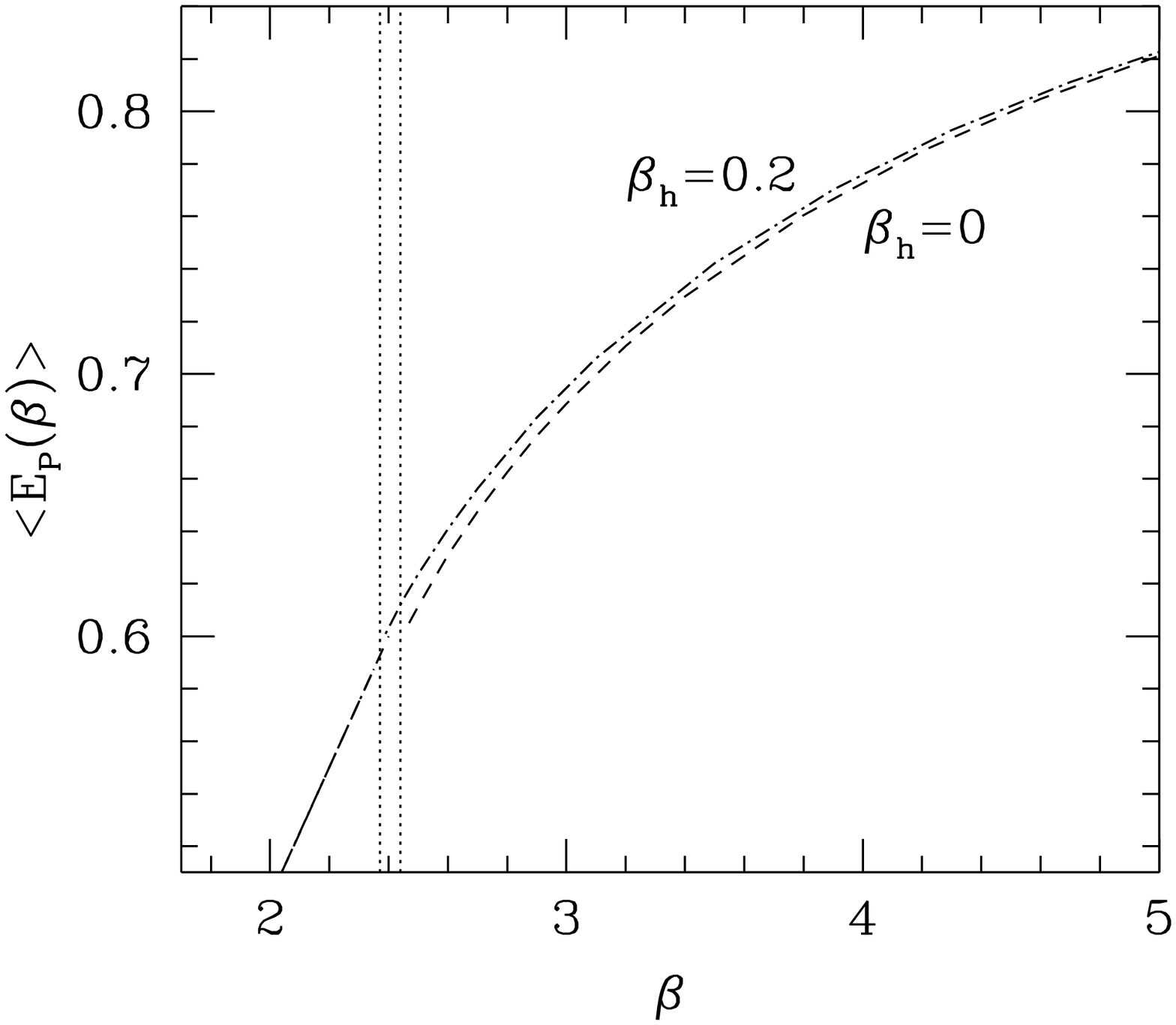,width=6cm,angle=0}
   \psfig{file=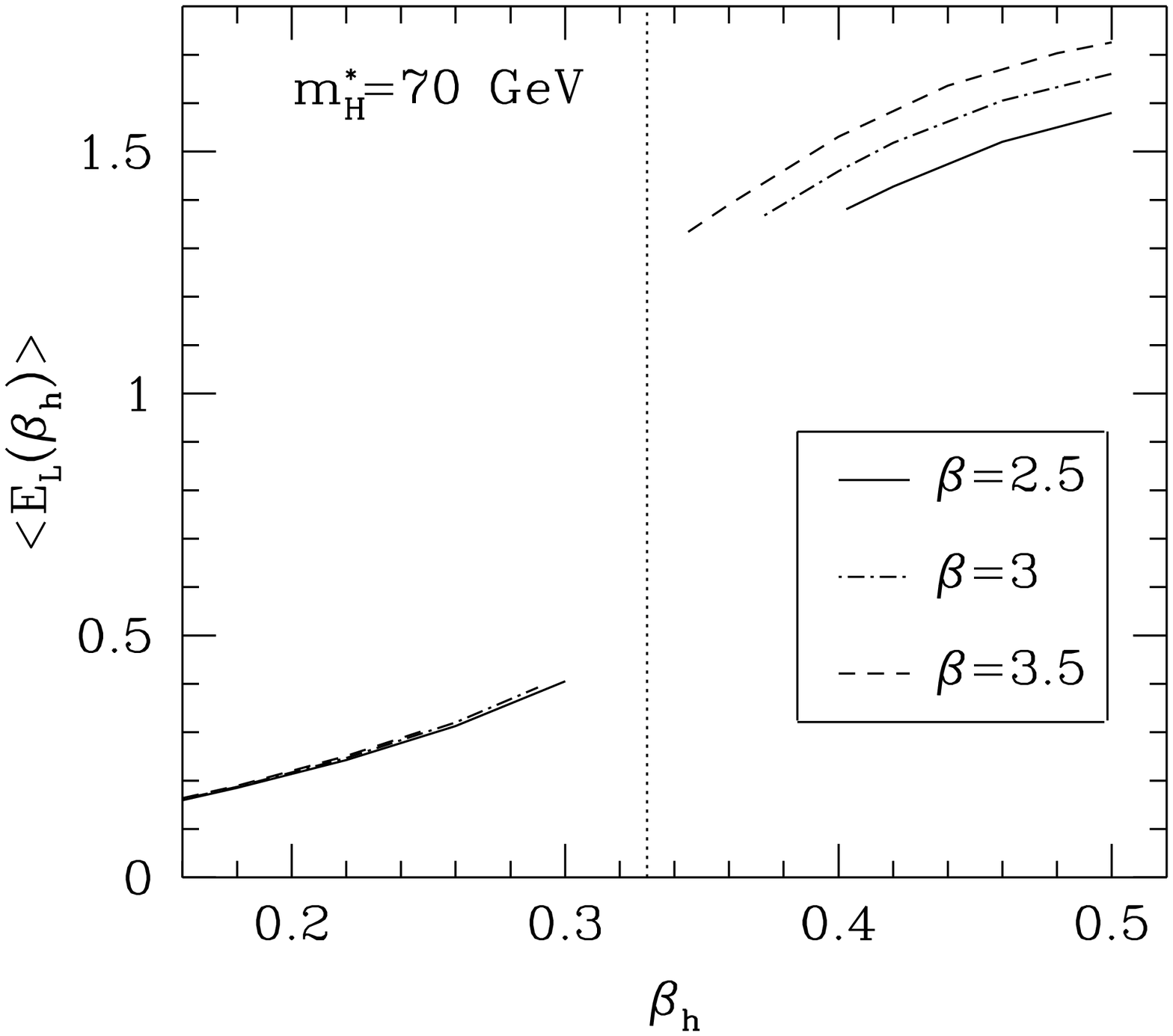,width=6cm,angle=0}%
   }
 \caption{On the left we plot $<E_P>_1$ versus $\beta$ for 
 ($\beta_h=0,\beta_R=0$)
 and also ($\beta_h=0.2,\beta_R=0.5$), a smooth transition 
 between PMS branches occurring at $\beta\approx 2.44$ and 2.37
 respectively.
 On the right we plot $<E_L>_1$ versus $\beta_h$ for a $4D$ Higgs mass
 $m_H^*=70GeV$, and three values of $\beta$. The vertical line is simply
 a rough guide to the position of the transition region
 $\beta_h=0.31-0.35$.} 
\end{figure} 

In Fig.~1 results in the transition/crossover regions 
for the average plaquette and hopping term calculated
to first order in $\de$ are shown. For each observable
tha parameter $J$ was fixed using (\ref{pms}). A PMS extremum solving
(\ref{pms}) exists for $<E_P>$
on both sides of the deconfinement crossover, with the strong
coupling solution corresponding to $J=0$ and a weak coupling PMS point
appearing only for $\beta>2.4$. Comparison with Monte Carlo
data available in $4D$ at $\beta_h=0$ indicates very good
agreement in the crossover region even at this low
order. For non-zero $\beta_h$ the transition shifts slightly to lower
$\beta$, again in agreement with Monte Carlo studies\cite{4d}.

For the hopping term we 
plot the results with parameters appropriate to an approximate
continuum $4D$ Higgs mass of $m_H^*=70$ GeV (see Ref. 2). 
Particular branches of the solution to (\ref{pms})
in each phase are lost as one approaches the transition region, leading
to difficulties in pinpointing the precise critical parameters for the
transition. It is 
expected that the branches will persist closer to the true transition point
as one evaluates the results to higher order in $\de$, and such a higher
order calculation is currently in progress.

\section*{Acknowledgements}
The financial support of T.S.E by the Royal Society, and 
A.R. by the Commonwealth Scholarship Commission is gratefully acknowledged.
This work was supported in part by the European Commission under 
the Human Capital and Mobility programme, contract number CHRX-CT94-0423.

\end{document}